\documentclass{PoS}
\usepackage{graphicx}
\usepackage{epsfig}
\usepackage[figuresright]{rotating}
\title{Outlook for Charged Higgs Physics}

\ShortTitle{Outlook}

\author{
{John~Ellis}
\thanks{\large CERN-PH-TH/2009-004}\\
        Theory Division, Physics Department, CERN, CH-1211 Geneva 23, Switzerland\\
        E-mail: \email{John.Ellis@cern.ch}}

\abstract{Almost all extensions of the Standard Model predict the existence of
charged Higgs bosons. This talk focuses on the minimal supersymmetric extension
of the Standard Model (MSSM), which is relatively predictive. The outlook for detecting
supersymmetric particles and Higgs bosons at the LHC are discussed, as are the
prospects for finding indirect effects of supersymmetric Higgs bosons at low energies,
e.g., in $K$ decays. The outlook for discovering observable effects of 
CP-violating supersymmetric phases at high energies or in $B$ decays is also mentioned.
}

\FullConference{Prospects for Charged Higgs Discovery at Colliders\\
		 September 16-19 2008\\
		 Uppsala, Sweden}

\begin{document}

\section{Beyond the Minimal Higgs Model}

Rare is the theorist who advocates the minimal Higgs model for
electroweak symmetry breaking (EWSB), based on a single elementary doublet of
Higgs fields. It is well-known that quantum corrections render the required small
mass scale $m_W \ll m_{GUT}$ or $m_P$ very unnatural, and this motivates many
non-minimal Higgs models, such as composite Higgs scenarios and supersymmetric
extensions of the Standard Model. Accordingly, (almost) all models of EWSB are 
non-minimal, featuring either one or more additional Higgs doublets, and possibly
other Higgs fields, either singlets or triplets of weak SU(2). The only scenarios
without charged Higgs bosons are those where the supplementary Higgs fields
are all singlets, but the preferred scenarios are those with extra Higgs doublets,
and hence charged Higgs bosons.
 
The minimal two-Higgs-doublet model (THDM) has five physical Higgs bosons: two
charged ($H^\pm$) and three neutral, two of which are scalars (the lighter $h$
and the heavier $H$)
and one pseudoscalar ($A$). The most studied type of THDM is the
minimal supersymmetric extension of the Standard Model (MSSM)~\cite{MSSM}, in which the
effective Higgs potential has specific restrictions, resulting from the specification in
supersymmetry of the quartic Higgs couplings in terms of the gauge couplings, and the 
pattern of supersymmetric radiative corrections. There was considerable discussion
of more general THDMs during this workshop, but in this talk I focus on variations of the MSSM,
since these are the most predictive. 

\section{Supersymmetric Higgs Models}

\begin{figure}
\begin{center}
\resizebox{0.8\textwidth}{!}{%
  \includegraphics{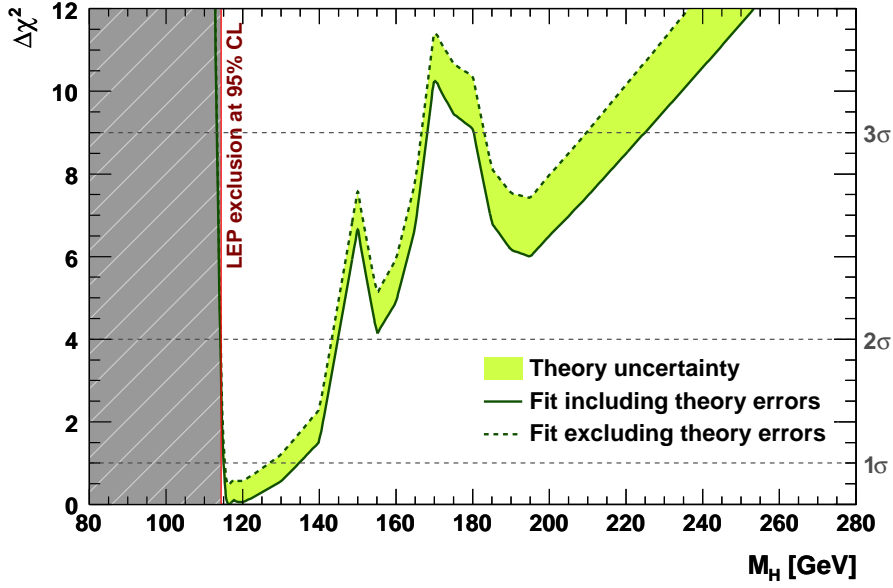}
}
\end{center}
\caption{\it The $\chi^2$ function for the mass of the Higgs boson in the
Standard Model~\protect\cite{Gfitter}, combining the negative results of direct searches at 
LEP~\protect\cite{LEPH} and the Tevatron collider~\protect\cite{TevH} with the indications 
from precision electroweak measurements\protect\cite{data}.
}
\label{fig:Gfitter}       
\end{figure}

This concentration on supersymmetry merits a few more words of motivation.
There are many reasons to like supersymmetry: its intrinsic beauty, its use in rendering
the hierarchy $m_W \ll m_{GUT}$ or $m_P$ more natural~\cite{hierarchy}, its help in promoting the
unification of the gauge couplings~\cite{GUTs}, the fact that it predicts a relatively light Higgs boson
weighing $< 150$~GeV~\cite{lightH}, as suggested by precision electroweak data~\cite{data}, 
the fact that it
provides a plausible candidate for the cold dark matter~\cite{DM}, and the fact that it is an (almost)
essential ingredient in string theory. Fig.~\ref{fig:Gfitter} displays the latest indications
concerning the Higgs mass~\cite{Gfitter}, showing the likelihood function obtained by combining
the negative results of Higgs searches at LEP and (more recently) the Tevatron with
the indications on the Higgs mass provided by precision electroweak data. Taken
together, they suggest strongly that $m_h \in (114, 140)$~GeV, in good agreement
with the prediction of supersymmetry for the mass of the lighter scalar Higgs boson.
We will see later what indications there may be on the possible masses of the
heavier supersymmetric Higgs bosons.

Enthused now by supersymmetry, we now focus on specific supersymmetric models. 
Within the MSSM framework, the properties of the Higgs sector are characterized
at the classical level by the supersymmetric coupling $\mu$ between the Higgs doublets
and the related bilinear soft supersymm- etry-breaking parameter $B$, by the
ratio of Higgs vacuum expectation values (VEVs), and by the supersymmetry-breaking
contributions $m_0$ to the masses of the two Higgs doublets. At the loop level,
there is also sensitivity to the supersymmetry-breaking contributions $m_0, m_{1/2}$ 
to the masses of the other MSSM particles, and to the soft trilinear supersymmetry-breaking
parameters $A$. It is often assumed that these parameters are universal at the GUT scale, 
a framework referred to as the constrained MSSM (CMSSM). This is not the same as 
minimal supergravity (mSUGRA), in which there are additional relations for the
gravitino mass: $m_{3/2} = m_0$, and between the bilinear and trilinear
supersymmetry-breaking parameters: $B = A Ð m_0$.

One may also consider scenarios with non-universal supersymmetry-breaking
scalar masses. Universality for the different sfermions with the same quantum numbers
e.g., $d, s$ squarks, is favoured by upper limits on flavour-changing neutral interactions
beyond the Standard Model~\cite{EN,BG}. These arguments do not extend to squarks with different
quantum numbers, or to squarks and sleptons, but these are also expected to be universal
in various GUT models, e.g., $m_{\tilde d_R} = m_{\tilde e_L}, m_{\tilde d_L} = m_{\tilde u_L}
 = m_{\tilde u_R} = m_{\tilde e_R}$ in SU(5), all squark and slepton masses equal in SO(10).
However, there are no arguments to forbid non-universal supersymmetry-breaking 
contributions to the masses of the Higgs masses. If one allows these to vary independently
of the common scalar masses $m_0$ of the squarks and sleptons, there are one or
more extra  parameters compared to the CMSSM, depending whether the Higgs masses
are assumed to be equal (NUHM1) or allowed to differ (NUHM2)~\cite{NUHM}.

In what follows, we will use the CMSSM, NUHM1 and NUHM2 as guides to the outlook
for charged Higgs physics.

\section{Constraints on Supersymmetry}

\begin{figure}[ht]
\resizebox{0.45\textwidth}{!}{
\includegraphics{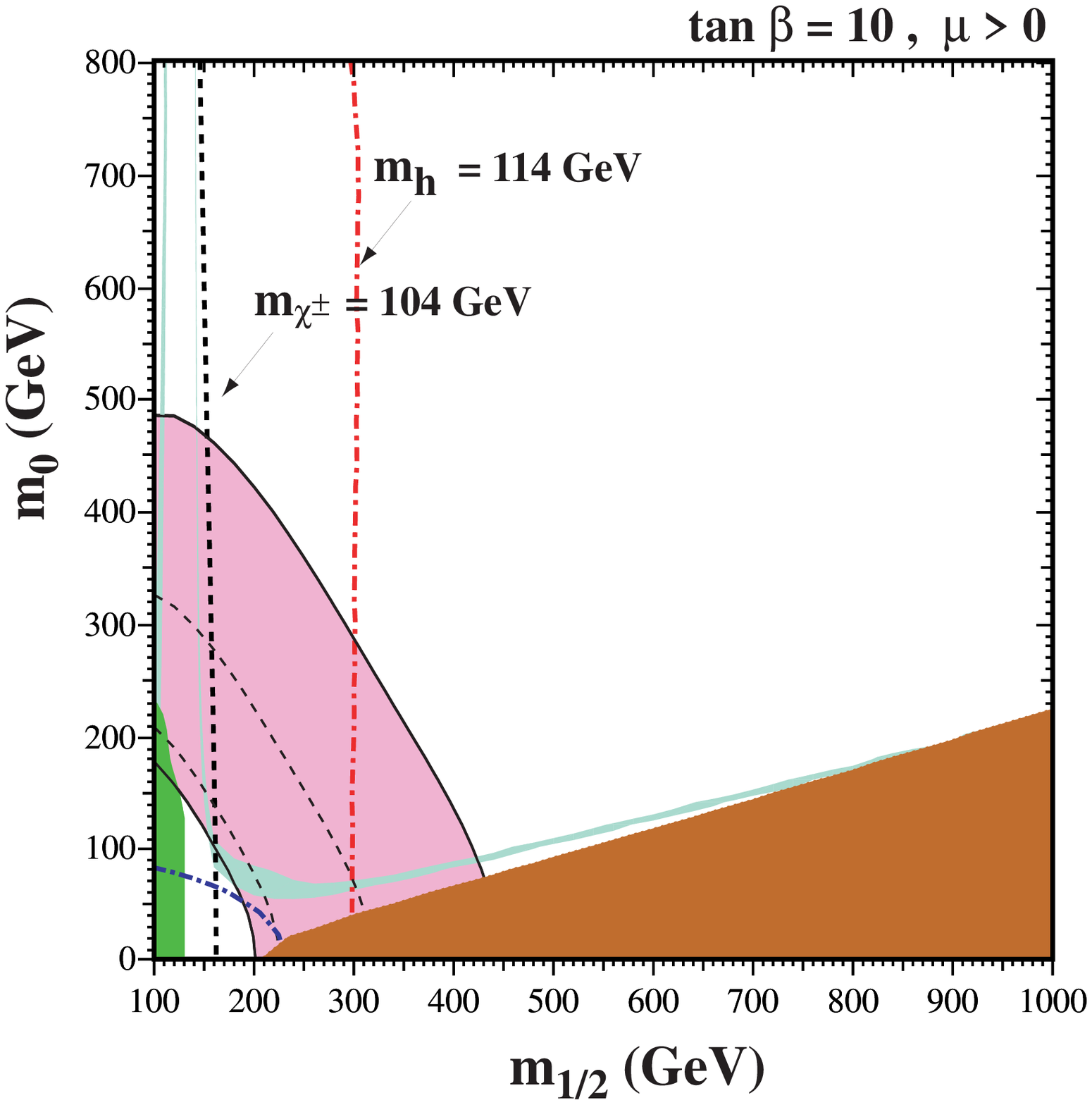}}
\resizebox{0.45\textwidth}{!}{
\includegraphics{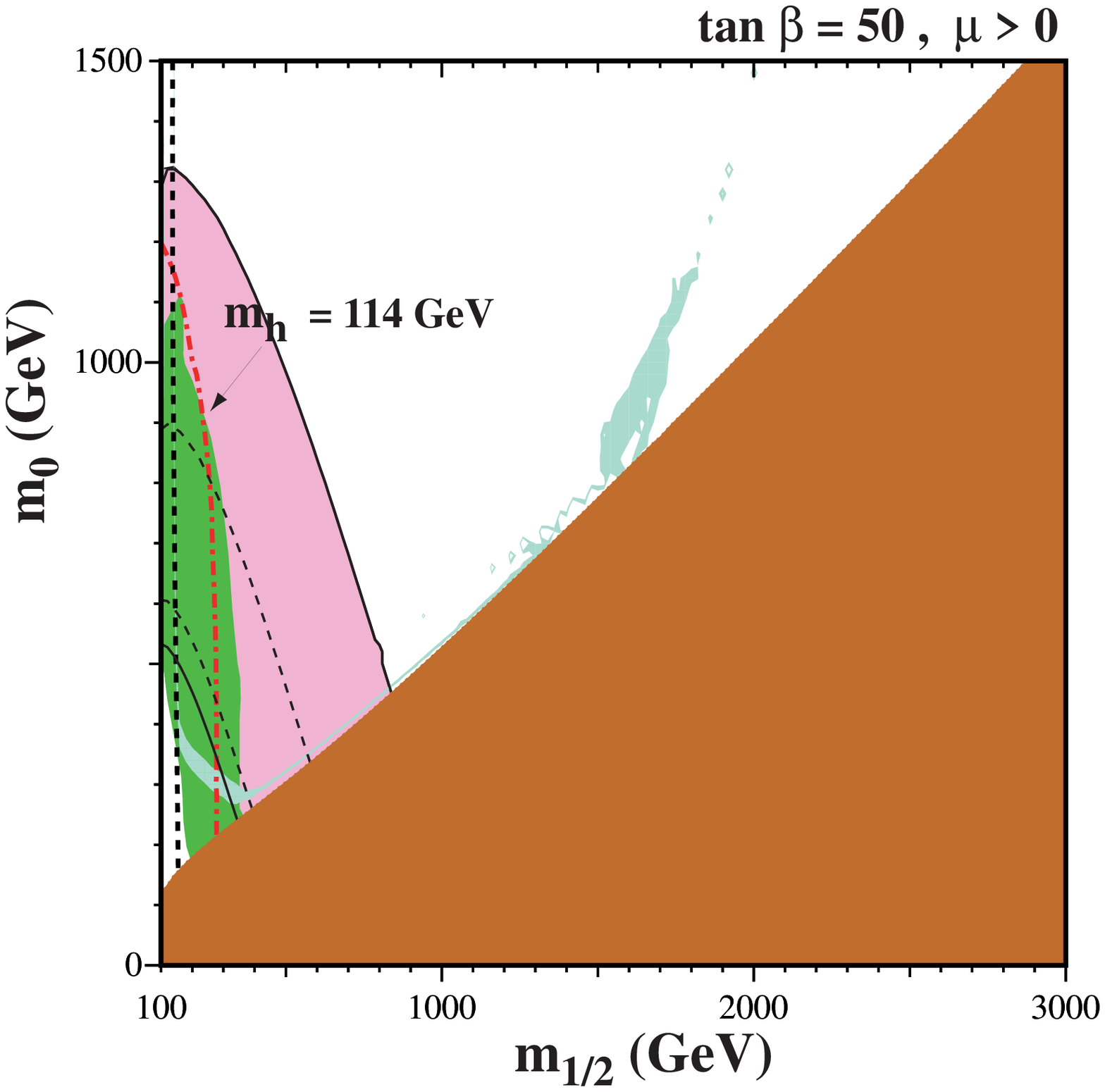}}
\caption{\label{fig:CMSSM}
{\it The $(m_{1/2}, m_0)$ planes for  (left) $\tan \beta = 10$ and (right) $\tan \beta = 50$,
assuming $\mu > 0$, $A_0 = 0$, $m_t = 175$~GeV and
$m_b(m_b)^{\overline {MS}}_{SM} = 4.25$~GeV~\protect\cite{EOSS}. The near-vertical (red)
dot-dashed lines are the contours for $m_h = 114$~GeV, and the near-vertical (black) dashed
line is the contour $m_{\chi^\pm} = 104$~GeV. Also
shown by the dot-dashed curve in the lower left is the region
excluded by the LEP bound $m_{\tilde e} > 99$ GeV. The medium (dark
green) shaded region is excluded by $b \to s
\gamma$, and the light (turquoise) shaded area is the cosmologically
preferred region. In the dark
(brick red) shaded region, the LSP is the charged ${\tilde \tau}_1$. The
region allowed by the E821 measurement of $a_\mu$ at the 2-$\sigma$
level, is shaded (pink) and bounded by solid black lines, with dashed
lines indicating the 1-$\sigma$ ranges.}}
\end{figure}

There are important constraints on supersymmetric models from the
absence of sparticles at LEP, which tell us that the
selectron and chargino weigh more than about 100~GeV~\cite{LEP}, and the Tevatron~\cite{Tevatron},
which tells us that squarks and gluinos weigh more than about 300~GeV.
There are also important indirect constraints from the LEP lower limit on the
Higgs mass of 114~GeV~\cite{LEPH}, and from the measured rate for $b \to s \gamma$~\cite{bsg}
(other $B$-decay constraints such the upper limit on $B_s \to \mu^+ \mu^-$
decay are also important at larger $\tan \beta$).
The density of astrophysical cold dark matter: $0.094 < \Omega_{CDM} h^2 < 0.124$~\cite{WMAP}
is also an important constraint. The most studied supersymmetric candidate for the
cold dark matter particle is the lightest neutralino $\chi$. Assuming this to be the
case, the relic density range tightly constrains one combination of the
supersymmetric model parameters, restricting them to a narrow strip in the sample CMSSM
$(m_{1/2}, m_0)$ plane shown in Fig.~\ref{fig:CMSSM}~\cite{EOSS}. We see that relatively heavy
sparticle masses would be compatible with the above constraints.

The most controversial constraint on the supersymmetric parameter
space is that imposed by the BNL measurement~\cite{BNL} of the
anomalous magnetic moment of the muon, $g_\mu -2$. There has been a
longstanding disagreement between the Standard Model predictions based on
low-energy $e^+ e^-$ data (which suggest a discrepancy exceeding 3 $\sigma$
that could easily be accommodated by supersymmetry~\cite{g-2th})
and on $\tau$ decay data (which do not exhibit any significant discrepancy). Recently, a new
preliminary $e^+ e^-$ measurement by the BABAR Collaboration using the
radiative-return technique has been announced, which agrees better with the $\tau$
decay data~\cite{Davier}. Therefore, the experimental situation currently remains unclear. 
Fig.~\ref{fig:CMSSM}
displays the region of the sample CMSSM $(m_{1/2}, m_0)$ plane that would be
favoured at the 1- and 2-$\sigma$ levels within the standard interpretation of the
$g_\mu -2$ measurement, based on $e^+ e^-$ data~\cite{EOSS}. If this constraint is valid, it
would suggest that sparticles may be relatively light.

\section{The LHC}

In the immediate future, the best outlook for charged Higgs physics lies with the LHC,
which started operation on Sept. 10th, shortly before this meeting. The CERN accelerator 
team was quickly able to circulate beams in both directions, to capture and bunch one beam
with the RF system, and to store it for tens of minutes~\cite{LHCstart}. 
These successes augur well for the
successful future operation of the LHC. All the experiments recorded
events, notably beam `splash' events in which the beam struck an upstream beam stop
and generated large numbers of particles that passed through the detectors. These
enabled timing, alignment and calibration issues to be resolved, and demonstrated that the 
detectors were operational and ready to take collisions.

There was therefore particularly strong disappointment on Sept. 19th when, during the
presentation of this talk, an electrical problem in the interconnect between two LHC magnets
caused a massive Helium leak and collateral damage that put a premature end to LHC operations 
for the rest of the year. At the time of writing, the source of the problem has been
understood, and it does not seem to be a major design problem~\cite{LHCstop}.
The damage is being made good, and precautions
are being taken to avoid such an incident in the future and to mitigate the consequences
of any similar incident, should one ever recur. The current plan is to complete the repairs
during the normal CERN winter accelerator shutdown period, then cool down the
repaired LHC sectors. Barring any unforeseen problems, it should be possible to restart 
operations in the Summer of 2009, aiming at collisions at 10~TeV in the centre of mass.

\section{How soon might Supersymmetry be Detected?}

\begin{figure}[ht]
\resizebox{0.5\textwidth}{!}{
{\begin{rotate}{90}~~~~~~~~~~~~~~~~~~~~~~~~~~~~~~
~~~~~~~~~~~~~~~~~~~{\Large $m_{1/2}$ [GeV]}\end{rotate}}
\includegraphics{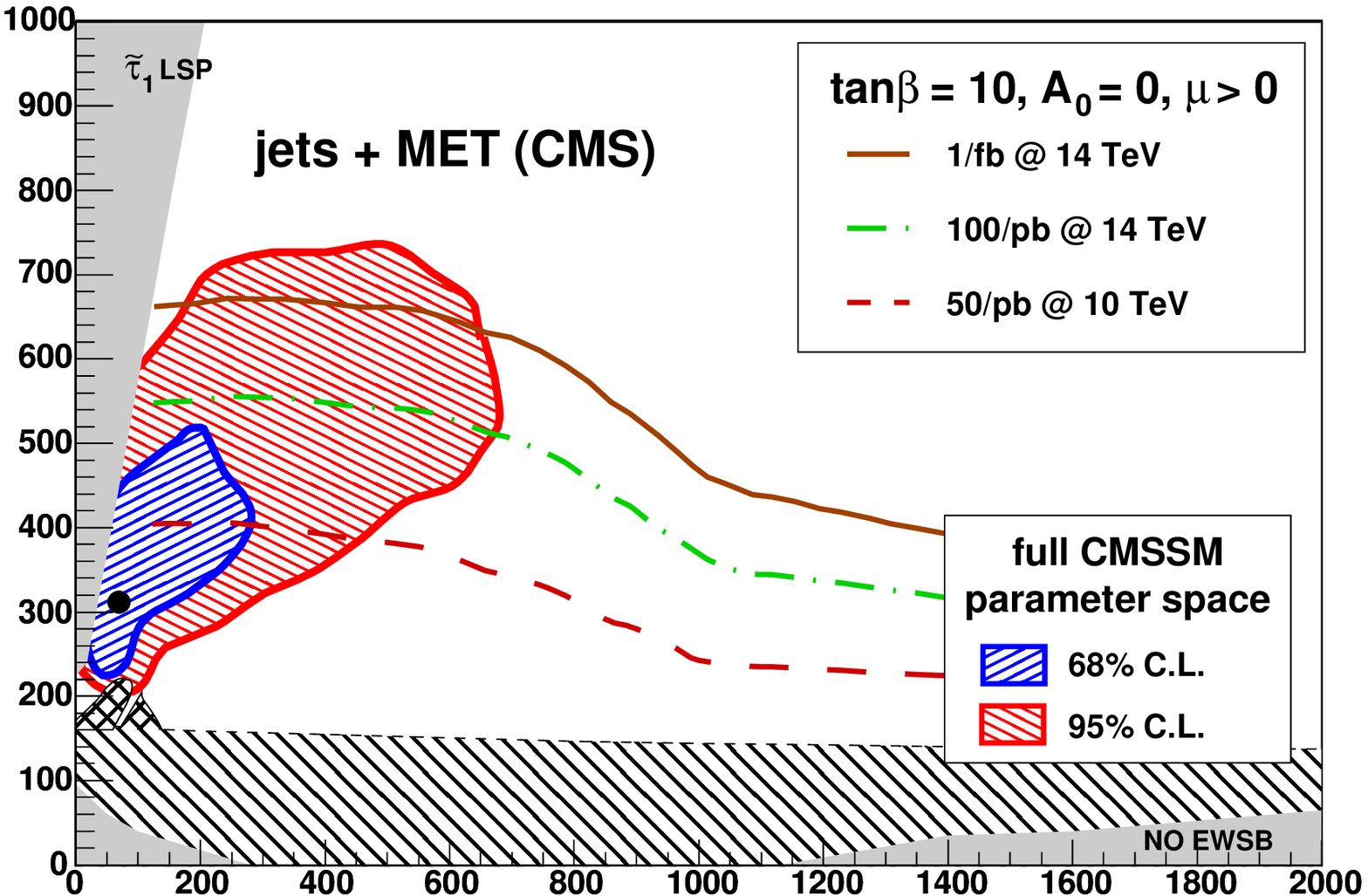}}
\resizebox{0.5\textwidth}{!}{
{\begin{rotate}{90}~~~~~~~~~~~~~~~~~~~~~~~~~~~~~~
~~~~~~~~~~~~~~~~~~~{\Large $m_{1/2}$ [GeV]}\end{rotate}}
\includegraphics{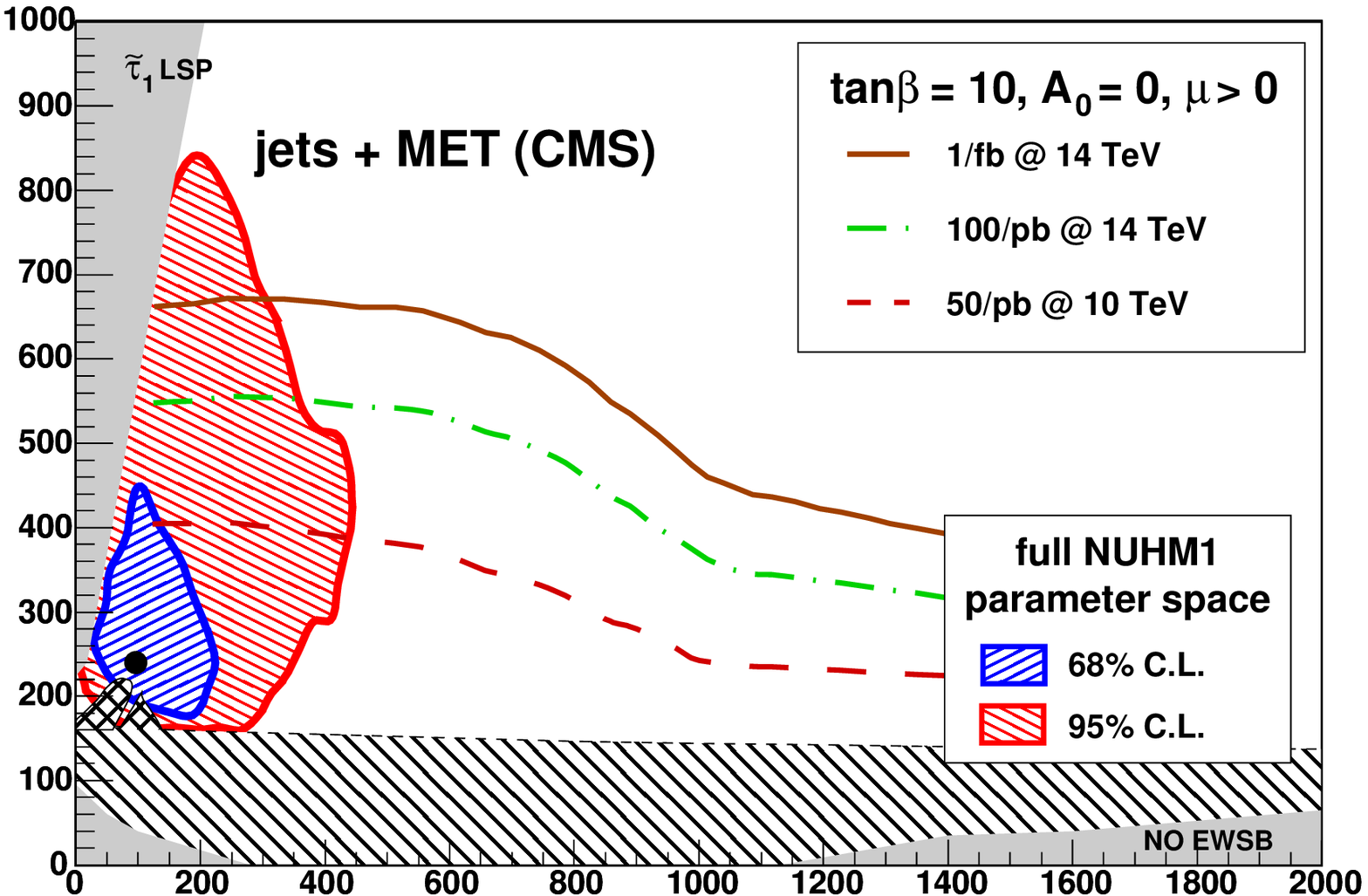}}
\vspace{-1cm}
\begin{flushright}
{\tiny $m_0$~[GeV]}
~~~~~~~~~~~~~~~~~~~~~~~~~~~~~~~~~~~~~~~~~~~~~~~~~~~~~~~~~~~~~~~~
{\tiny $m_0$~[GeV]}
~~~~~~~~~~~~~~~~~
\end{flushright}
\caption{\label{fig:master1}
{\it The $(m_0, m_{/2})$ planes for  (left) the CMSSM and (right) the NUHM1~\protect\cite{Master},
displaying the best-fit points (black dots), the 68\% C.L. regions (blue hatching), the
95\% C.L. regions (red hatching) and the region excluded by LEP~\protect\cite{LEP} 
and Tevatron~\protect\cite{Tevatron} searches (black hatching).
Also shown are the 5-$\sigma$ supersymmetry discovery reaches at the LHC assuming 
50/pb of data at 10~TeV, 100/pb at 14~TeV, and 1/fb at 14~TeV, as estimated for the indicated values
of the supersymmetric model parameters.
}}
\end{figure}

A likelihood analysis of the CMSSM and the NUHM1,
implementing all the constraints mentioned previously, including the controversial
$g_\mu - 2$ constraint, was recently performed in~\cite{Master}.
Fig.~\ref{fig:master1}(a) displays in the CMSSM $(m_{1/2}, m_0)$ plane the best-fit
point (black dot), and the regions of supersymmetric parameter space favoured at the 
68 and 95\% confidence levels (blue and red hatched regions, respectively). Also shown
are the regions of the $(m_{1/2}, m_0)$ plane excluded by previous LEP and Tevatron
searches (black hatched regions), and the regions where the LHC could discover 
supersymmetry at the 5-$\sigma$ level with the indicated amounts of luminosity:
50/pb at 10~TeV in the centre of mass, and 100/pb or 1/fb at the design energy of
14~TeV in the centre of mass. We see that the best-fit point could be discovered with just
50/pb at 10~TeV, that 100/pb at 14~TeV would suffice to discover supersymmetry in the
68\% confidence-level (C.L.) region, and that 1/fb at 14~TeV would cover almost all the
95\% C.L. region in the CMSSM. Fig.~\ref{fig:master1}(b) displays the
corresponding analysis for the NUHM1, which reaches rather similar conclusions,
though the favoured regions extend to somewhat larger values of $m_{1/2}$.
It should be emphasized, though, that these conclusions depend crucially on the
implementation of the questionable $g_\mu - 2$ constraint.

\begin{figure}[ht]
\resizebox{0.5\textwidth}{!}{
{\begin{rotate}{90}~~~~~~~~~~~~~~~~~~~~~~~~~~~~~~
~~~~~~~~~~~~~~~~~~~{\Large $m_{1/2}$ [GeV]}\end{rotate}}
\includegraphics{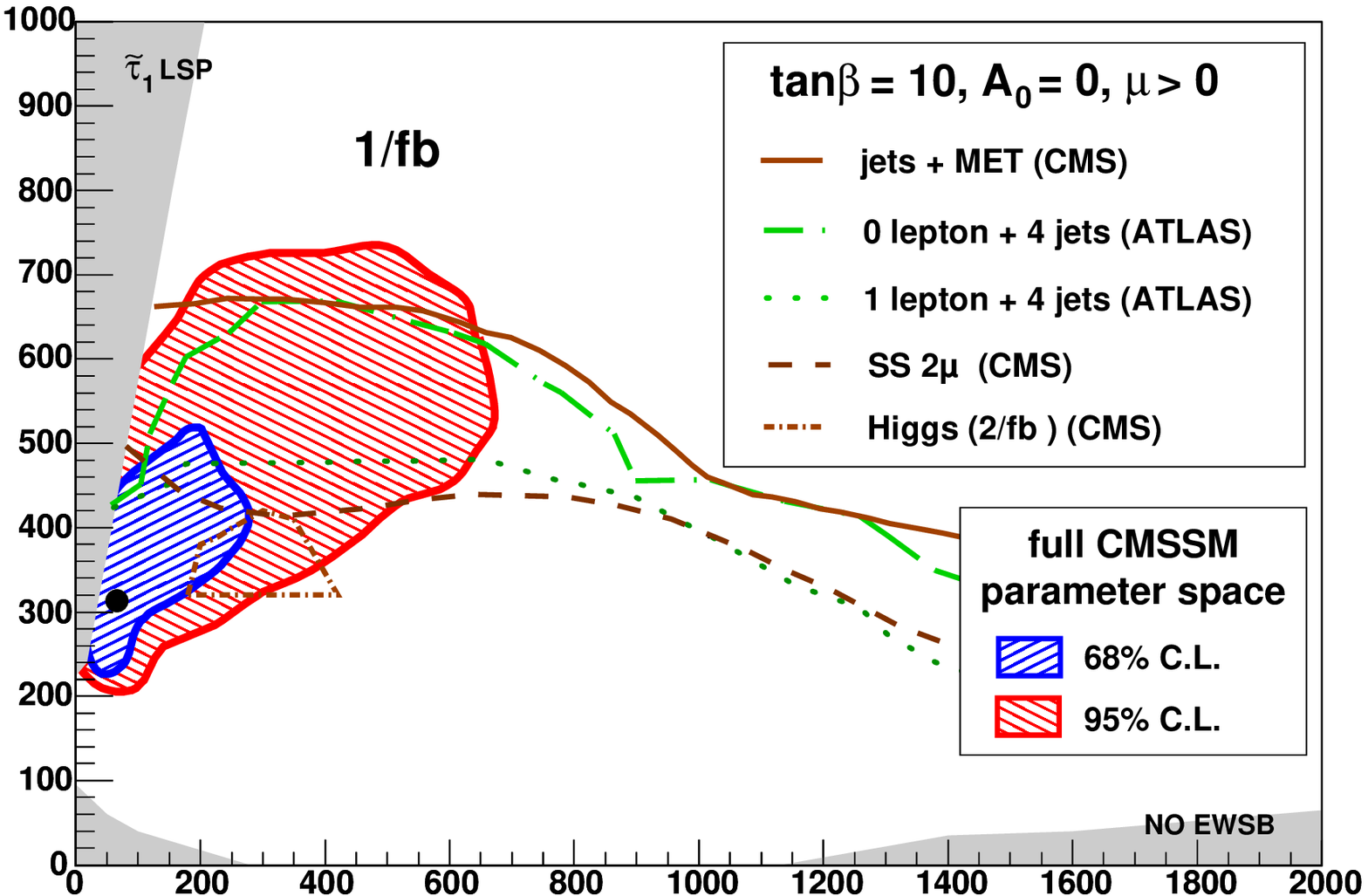}}
\resizebox{0.5\textwidth}{!}{
{\begin{rotate}{90}~~~~~~~~~~~~~~~~~~~~~~~~~~~~~~
~~~~~~~~~~~~~~~~~~~{\Large $m_{1/2}$ [GeV]}\end{rotate}}
\includegraphics{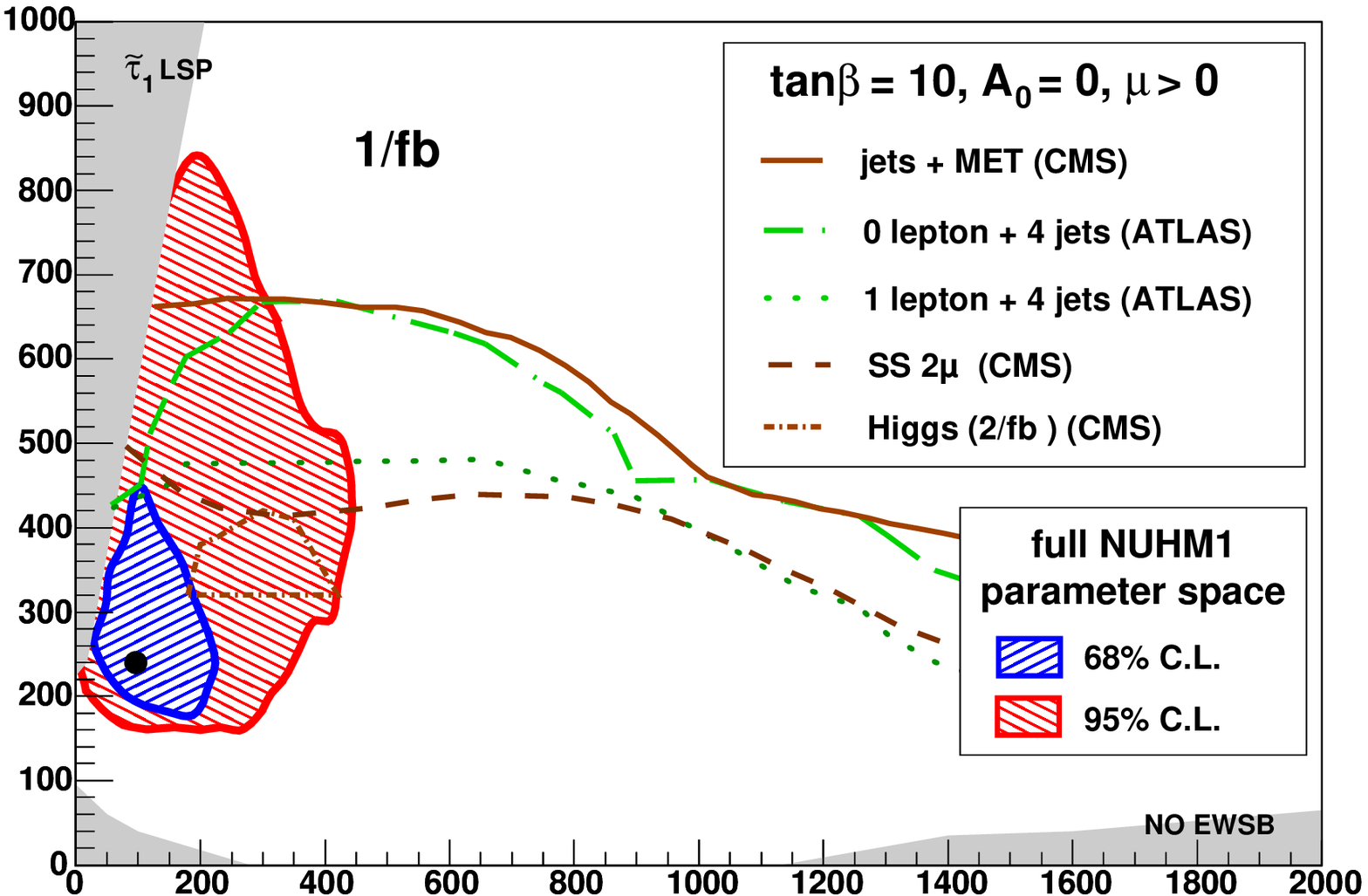}}
\vspace{-1cm}
\begin{flushright}
{\tiny $m_0$~[GeV]}
~~~~~~~~~~~~~~~~~~~~~~~~~~~~~~~~~~~~~~~~~~~~~~~~~~~~~~~~~~~~~~~~
{\tiny $m_0$~[GeV]}
~~~~~~~~~~~~~~~~~
\end{flushright}
\caption{\label{fig:master2}
{\it The same $(m_0, m_{/2})$ planes for  (left) the CMSSM and (right) the 
NUHM1 as in Fig.~\protect\ref{fig:master1}~\protect\cite{Master}, this time
displaying the LHC discovery reaches in several different channels with 1/fb at 14~TeV,
and the light Higgs discovery potential in supersymmetric cascades with 2/fb at 14~TeV, 
as estimated for the indicated values of the supersymmetric model parameters.
}}
\end{figure}

As shown in Fig.~\ref{fig:master2}, there are several channels in which supersymmetry
could be detected at the LHC, notably including various jet and lepton channels
accompanied by the classic missing-energy signal of the escaping dark matter
neutralinos. We also see that searches for the lightest supersymmetric Higgs boson
in supersymmetric decays could also play an interesting role, though perhaps not in the
most favoured regions of parameter space.

\begin{figure}[ht]
\resizebox{0.5\textwidth}{!}{
\includegraphics{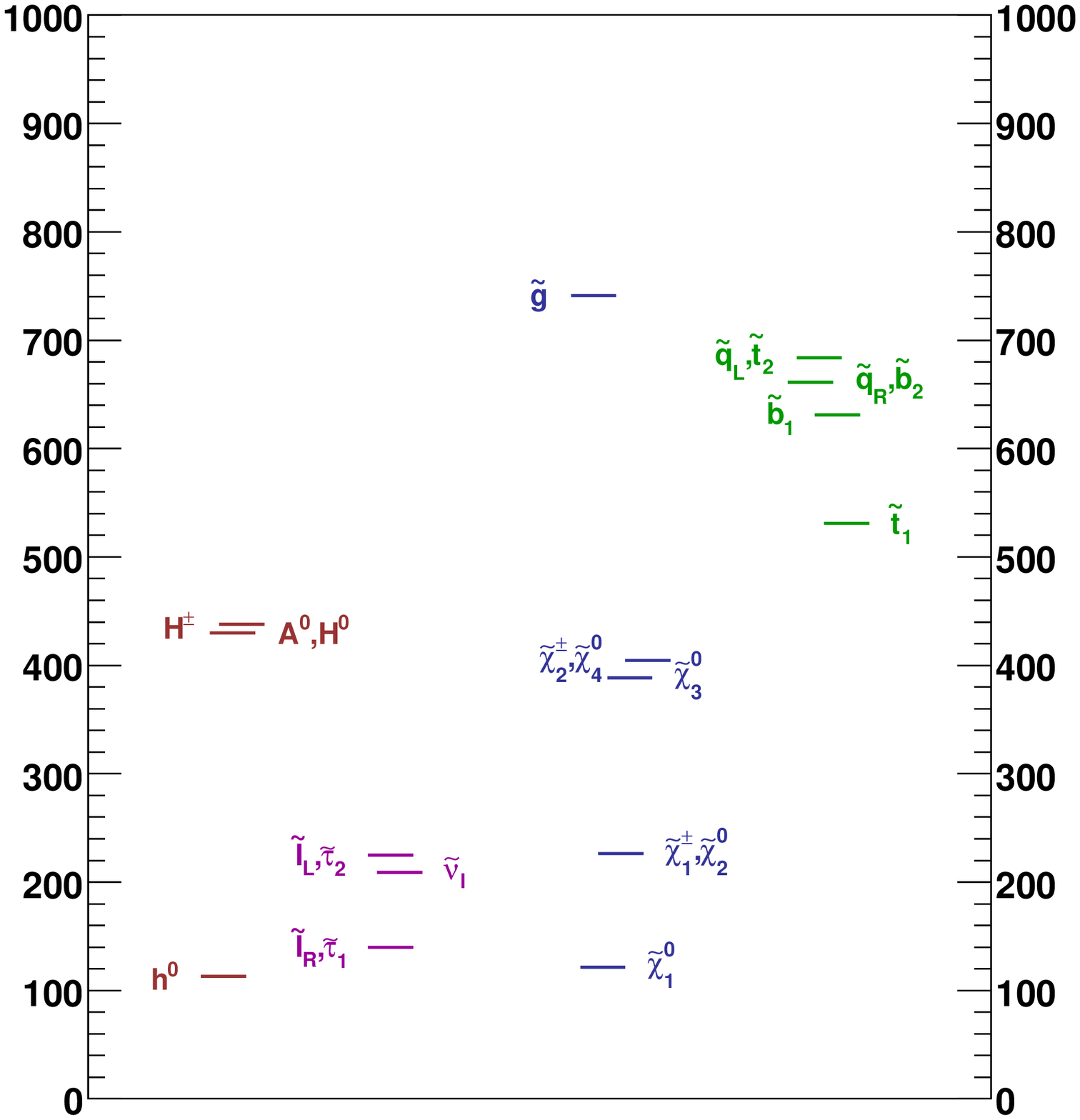}}
\resizebox{0.5\textwidth}{!}{
\includegraphics{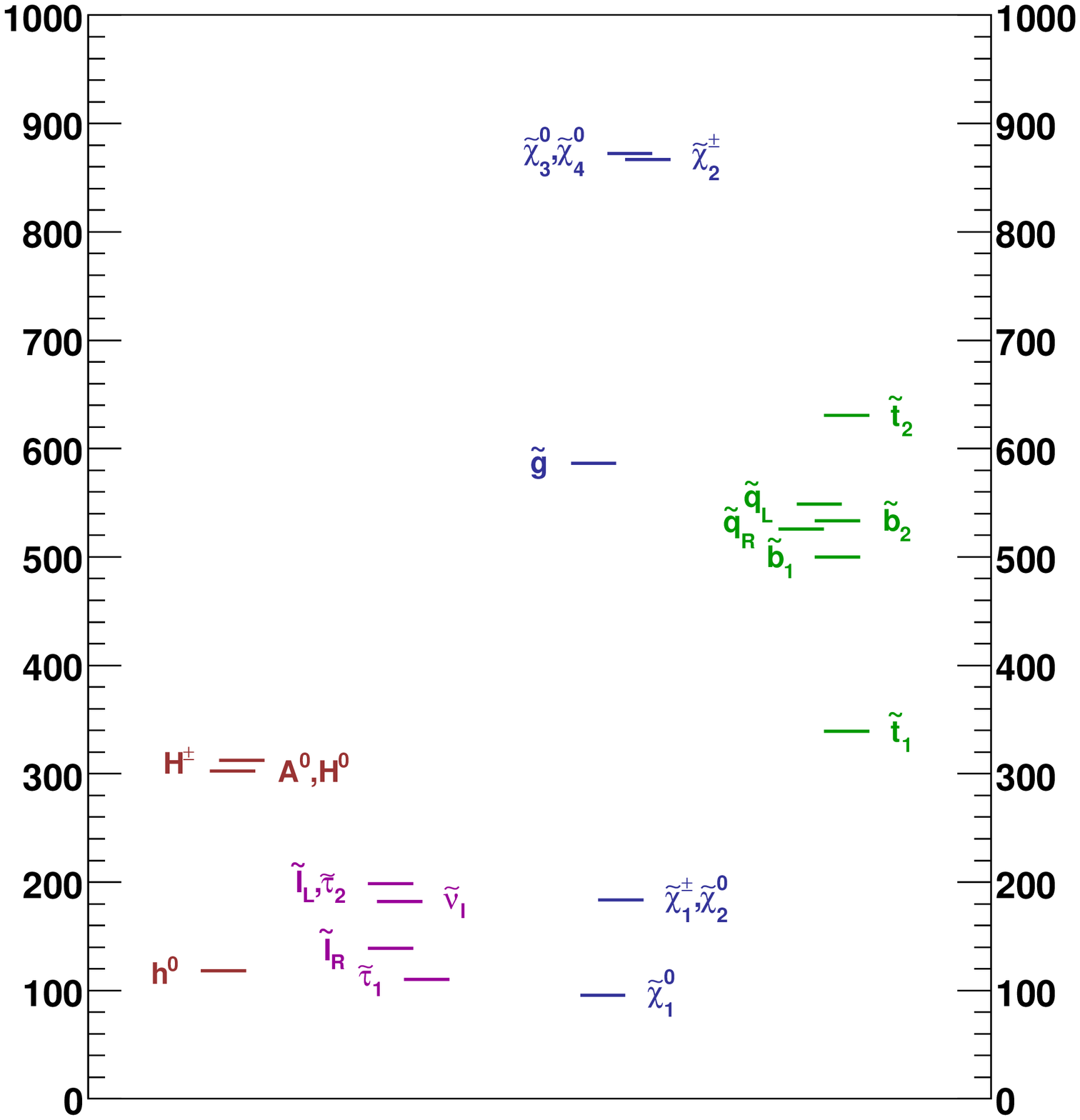}}
\caption{\label{fig:master3}
{\it The supersymmetric spectra at the best-fit points for (left) the CMSSM and
(right) the NUHM1~\protect\cite{Master}. Note that the heavier Higgs bosons, squarks 
sleptons, gluinos and lowest-lying inos are somewhat lighter in the NUHM1, whereas the
highest-lying inos are somewhat lighter in the CMSSM.}}
\end{figure}

The best-fit spectra in the CMSSM and NUHM1 are shown in Fig.~\ref{fig:master3}~\cite{Master}. 
We see that the favoured slepton, squark, gluino and lighter neutralino and chargino
masses are quite similar in the two scenarios, though somewhat lighter
in the NUHM1 because the best-fit value of $m_{1/2}$ is slightly lower. On the other hand,
the heavier neutralinos and chargino are significantly heavier in the NUHM1, and the
heavier Higgs bosons are significantly lighter. We will discuss later how this might affect
the prospects for discovering the heavier MSSM Higgs bosons at the LHC.

As can be seen in Fig.~\ref{fig:master2}, one of the most promising channels for
discovering a supersymmetric signal is in the same-sign dilepton signal (SS). An
early measurement of the edge of the dilepton mass spectrum might constrain
significantly the supersymmetric parameter space~\cite{Master},
and thereby refine the prospects for discovering the heavier Higgs bosons at the LHC
within various different models.

\section{Will the heavier Supersymmetric Higgs Bosons be discovered at the LHC?}

\begin{figure}[htb!]
\begin{center}
\includegraphics[width=0.45\textwidth]{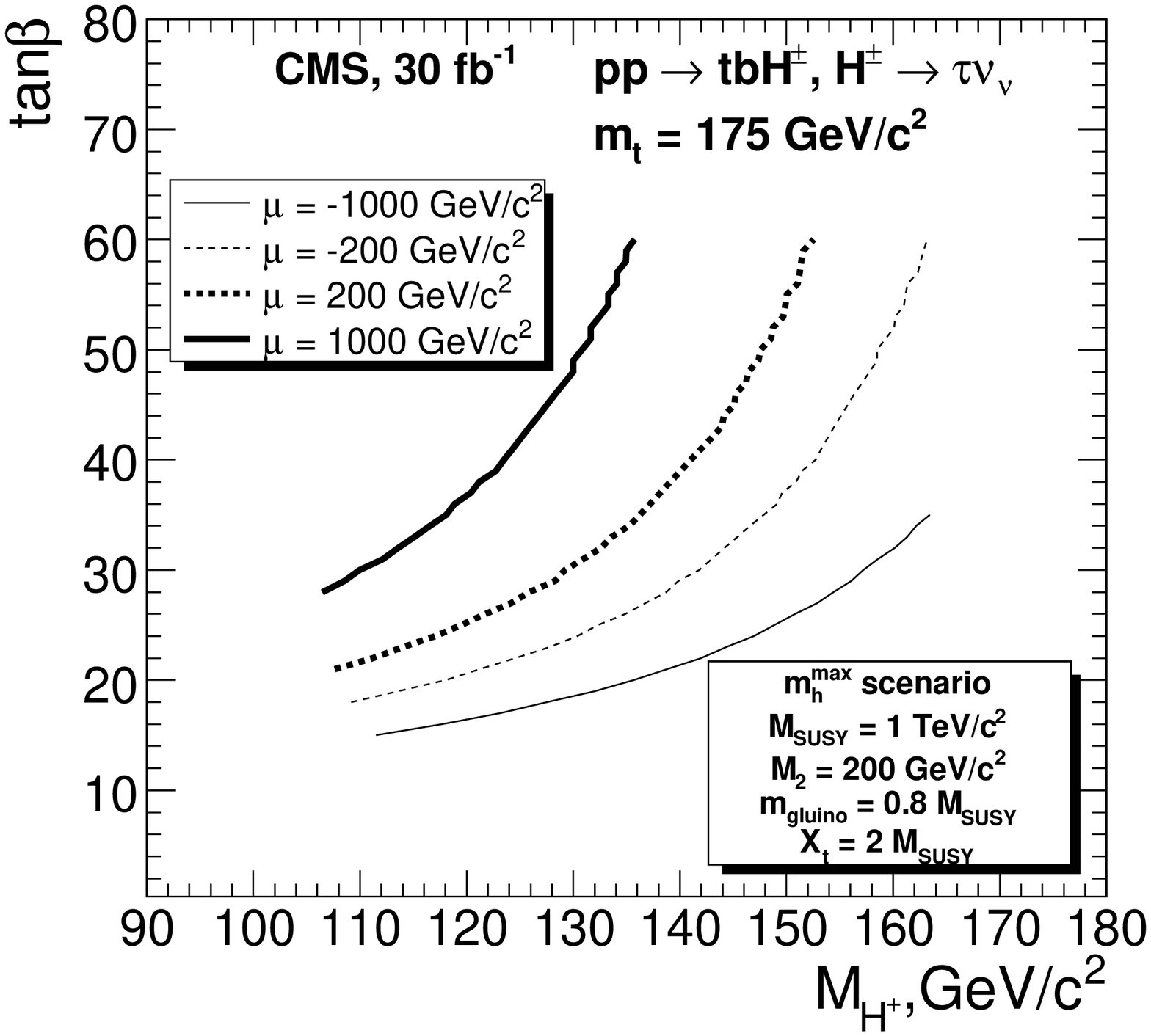}\hspace{1em}
\includegraphics[width=0.45\textwidth]{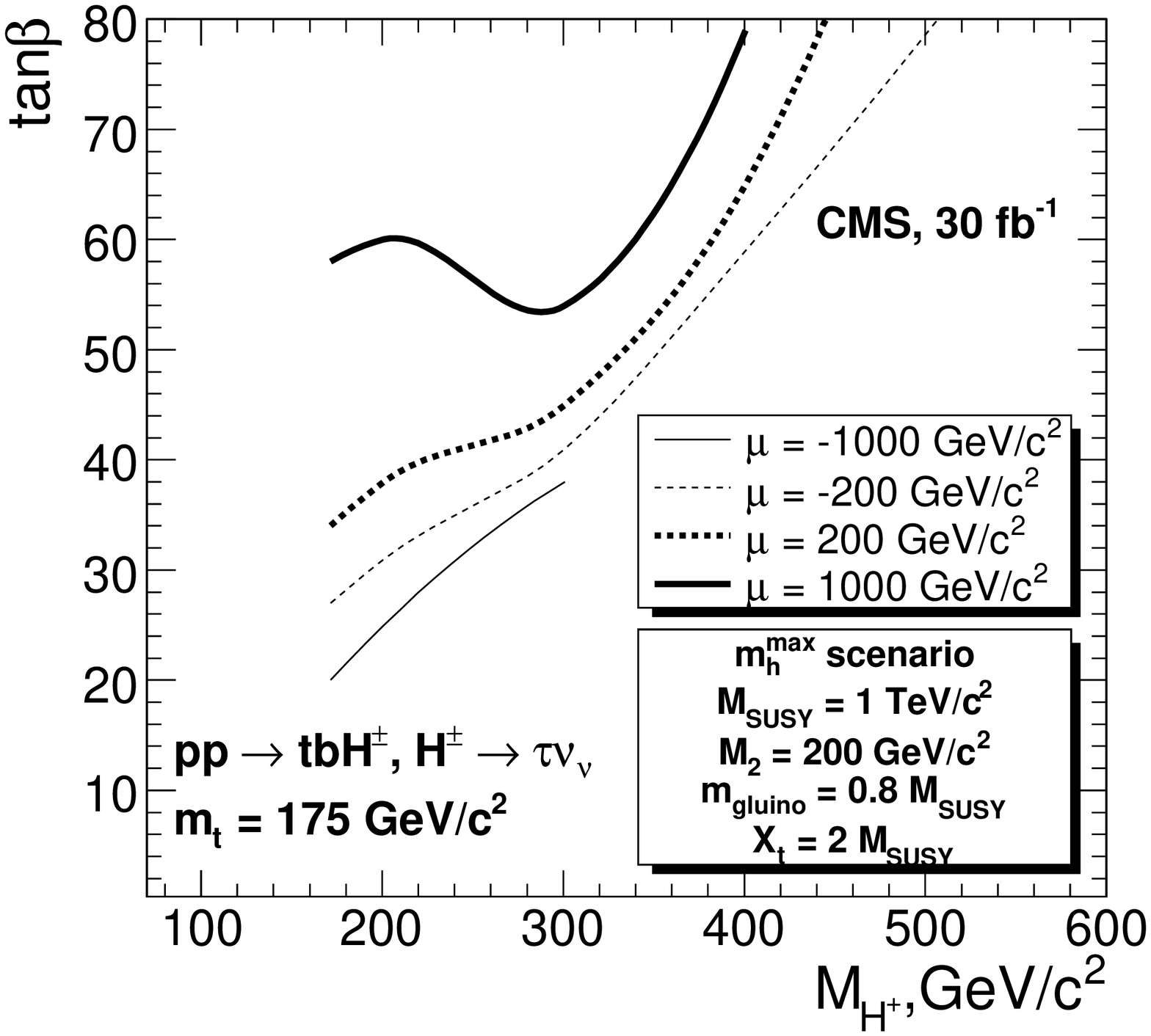}
 \caption{%
\it Discovery reach for (left) a light and (right) a heavy charged Higgs boson in the 
$(m_{H^\pm}, \tan \beta)$ plane for a maximal Higgs mixing scenario,
in CMS with 30/fb~\protect\cite{Hashemi}.
}
\label{fig:CMS}
\end{center}
\end{figure}

There have been many studies of the prospects for discovering the heavier
Higgs bosons at the LHC, usually presented as surveys of some $(m_A, \tan \beta)$
plane for fixed values of the other MSSM parameters~\cite{heavyHLHC}. 
Early studies indicated that the heavier Higgs bosons
might be detectable with 300/fb of data, no matter what the value of $\tan \beta$ if 
$m_A < 200$~GeV, and for $\tan \beta > 10$ if $m_A < 600$~GeV. The most
promising search channels seem to be $H^\pm \to \tau \nu$ or $t {\bar b}$ and
$H, A \to \mu \mu$ and $\tau \tau$. However, these analyses needed to be
validated with the latest and best available simulations of ATLAS~\cite{ATLAS} and CMS~\cite{CMS}.
Some updated sensitivities to charged Higgs bosons estimated assuming 30/fb of 
data are shown in Fig.~\ref{fig:CMS}~\cite{Hashemi}. We see that varying $\mu$ can have a significant
effect on the regions of the $(m_A, \tan \beta)$ plane that are accessible, and this is also
true of the other MSSM parameters displayed in the legends.

\begin{figure}
\begin{center}
\resizebox{0.6\textwidth}{!}{%
  \includegraphics{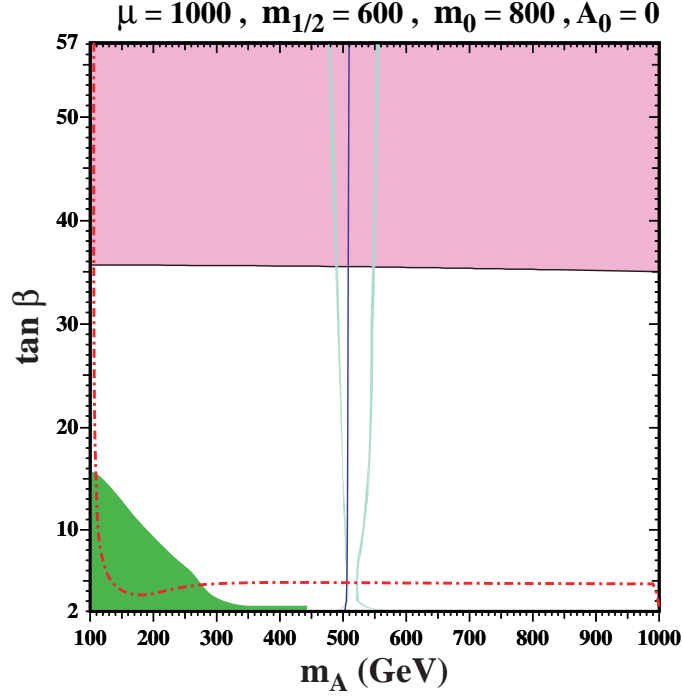}
}
\end{center}
\caption{\it The $(m_A, \tan \beta)$ plane in the NUHM2 with the indicated choices
of $\mu, m_{1/2}, m_0$ and $A_0$~\protect\cite{EHHOW}. The region excluded by $b \to s 
\gamma$ is shaded dark (green), and that favoured by $g_\mu - 2$ is shaded light (pink). The
near-vertical strips with the neutralino relic density in the range favoured by astrophysics are
shaded light (turquoise): they lie on either side of the (blue) line where $m_\chi = m_A/2$.
The LEP Higgs constraint is shown as a (red) dot-dashed line.}
\label{fig:EHHOW}       
\end{figure}

These and some previous analyses were made with specific choices of the low-energy 
MSSM parameters that are not necessarily possible within the CMSSM or NUHM1(2) frameworks. 
Nor do these analyses necessarily take into account all the phenomenological constraints
used in the previous analysis, specifically the cold dark matter constraint~\cite{EHHOW}. This point is
demonstrated in Fig.~\ref{fig:EHHOW}, where we see explicitly that most of
a sample $(m_A, \tan \beta)$ plane for certain {\it fixed} values of $\mu, m_{1/2}$ and $m_0$ is
compatible with $m_h$ and $b \to s \gamma$, but not necessarily with $g_\mu - 2$. More
importantly, we see that only two near-vertical, very narrow strips of the sample 
$(m_A, \tan \beta)$ plane on either side of the (blue) line where $m_\chi = m_A/2$
are compatible with the cold dark matter constraint. The band
between the two strips is under-dense, which might be permissible if there is some
additional source of cold dark matter. On the other hand, the exterior regions are 
{\it over-dense} and hence forbidden. However, one can construct an acceptable
$(m_A, \tan \beta)$ plane in the NUHM2 by {\it varying} $m_{1/2}$ with $m_A$, 
so as to maintain the appropriate relic density by keeping $m_\chi$ slightly 
$\ne m_A/2$~\cite{EHHOW}, e.g.,
\begin{equation}
m_{1/2} \sim \frac{9}{8} m_A \; \; {\rm for} \; \; m_0 = 800~{\rm GeV}, \mu = 1000~{\rm GeV}. 
\label{plane1}
\end{equation}
Acceptable NUHM2 planes can also be obtained~\cite{EHHOW} by varying $\mu$ with $m_A$, e.g.,
\begin{equation}
\mu \sim 250~{\rm to}~400~{\rm GeV} \; \; {\rm for} \; \; m_{1/2} = 500~{\rm GeV}, m_0 = 1000~{\rm GeV}. 
\label{plane2}
\end{equation}
These have been used to make global fits to the electroweak and $B$-decay observables
with the supersymmetric model parameters varying across these $(m_A, \tan \beta)$ planes, treating
the relic neutralino density as a fixed constraint. These serve as a more suitable NUHM2 framework
for assessing the outlook for heavy Higgs boson searches at the LHC and elsewhere.

\begin{figure}[htb!]
\vspace{10mm}
\begin{center} 
\includegraphics[width=.49\textwidth]{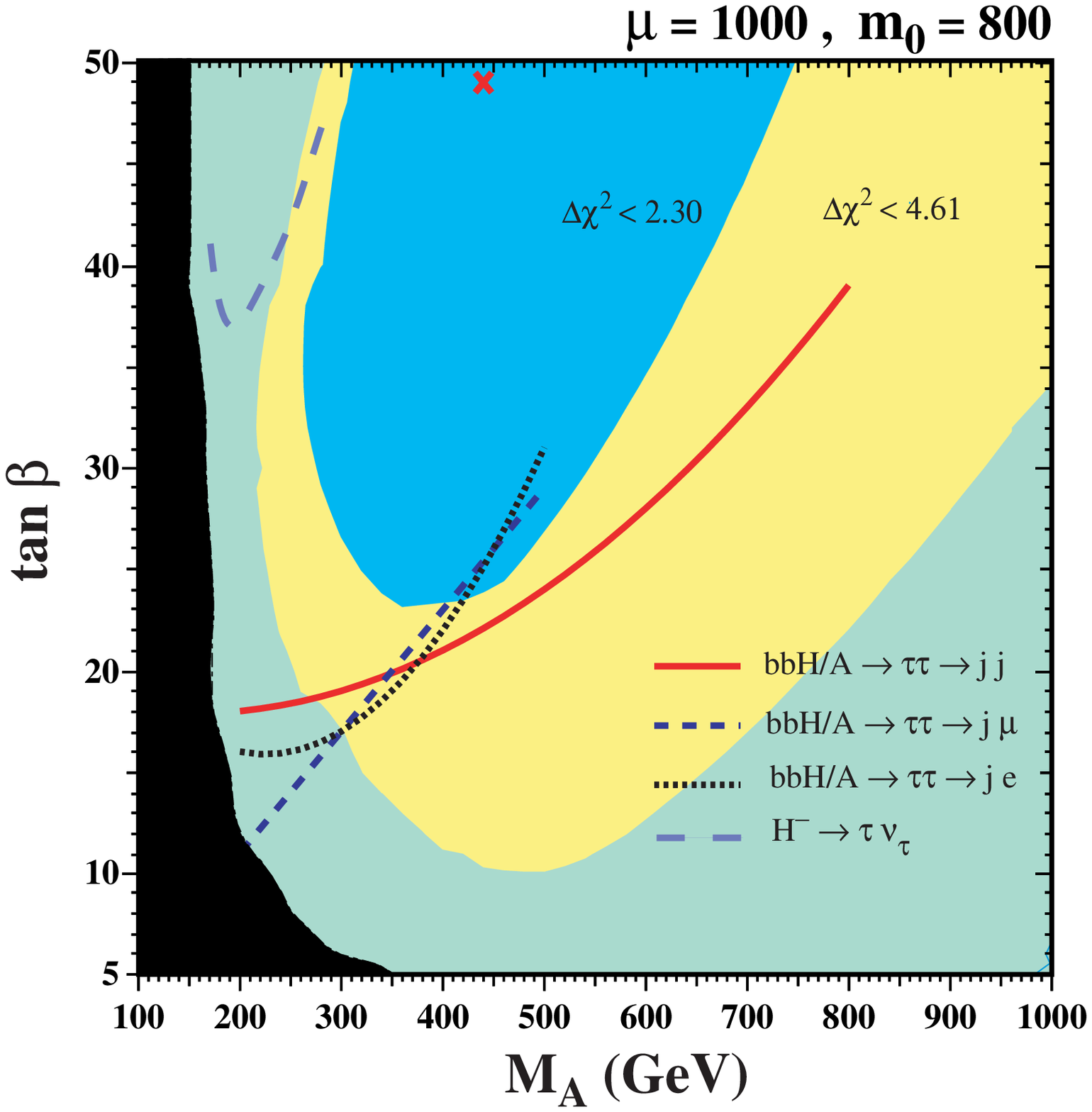}
\includegraphics[width=.49\textwidth]{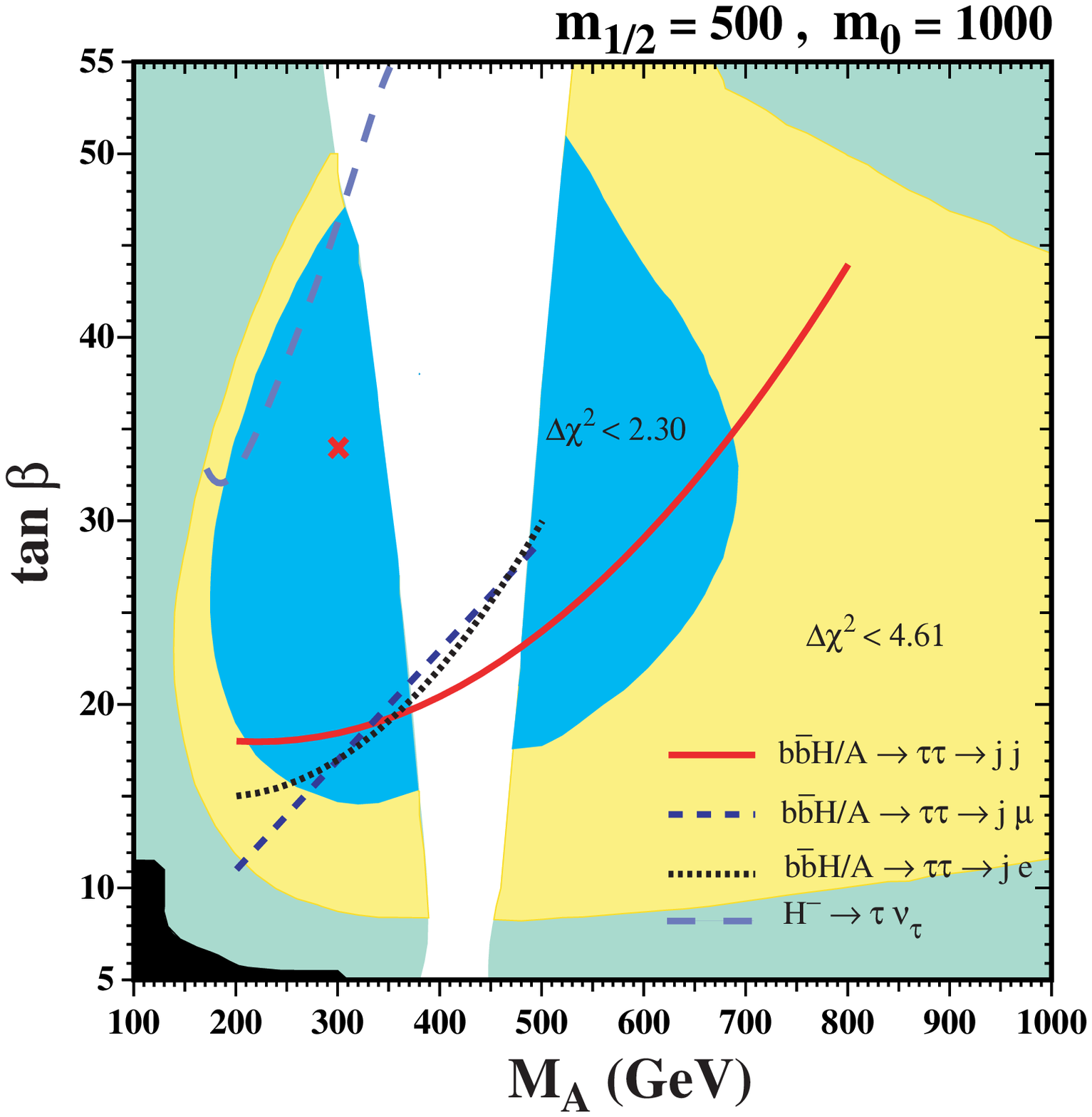}
\includegraphics[width=.49\textwidth]{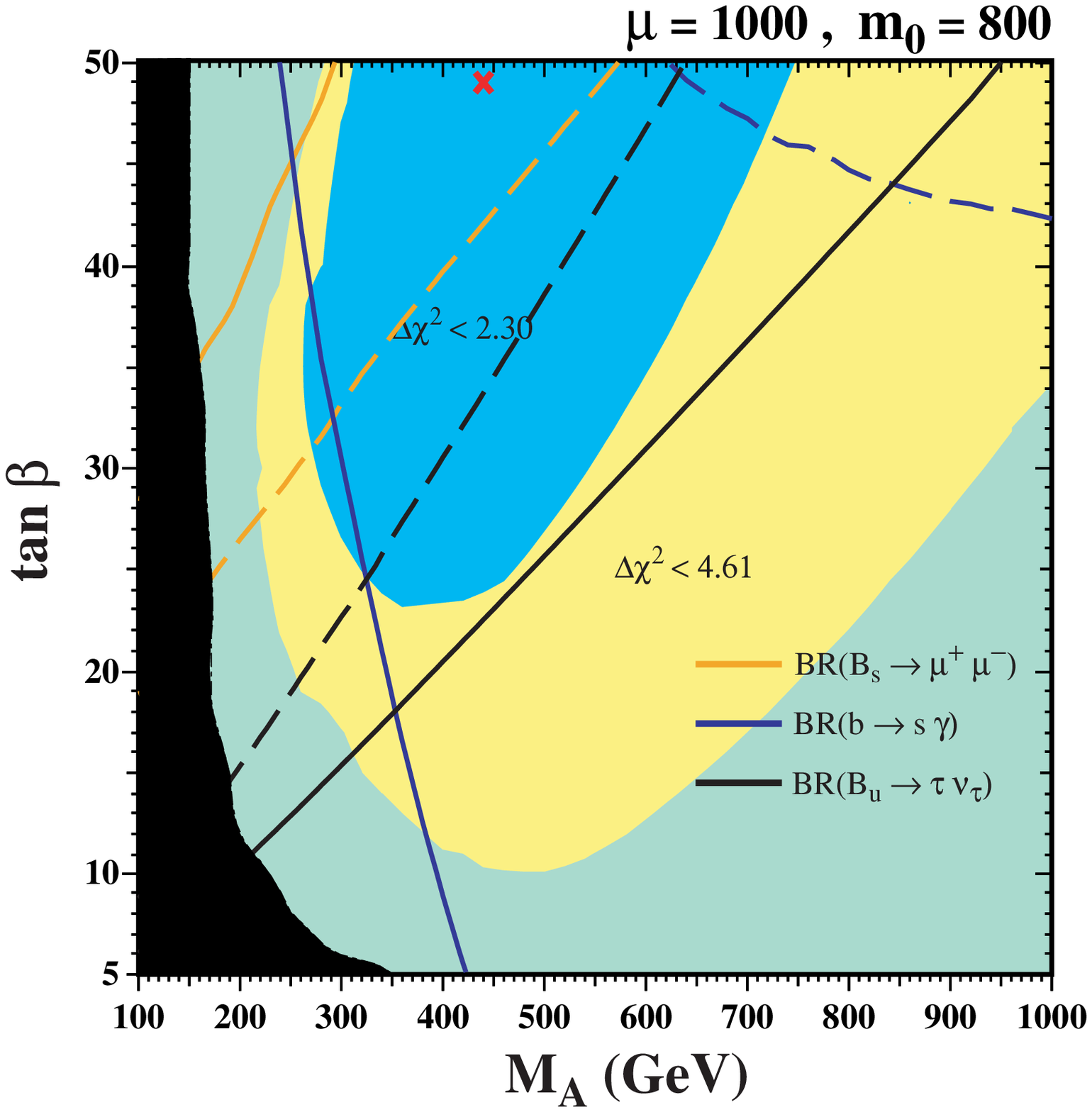}
\includegraphics[width=.49\textwidth]{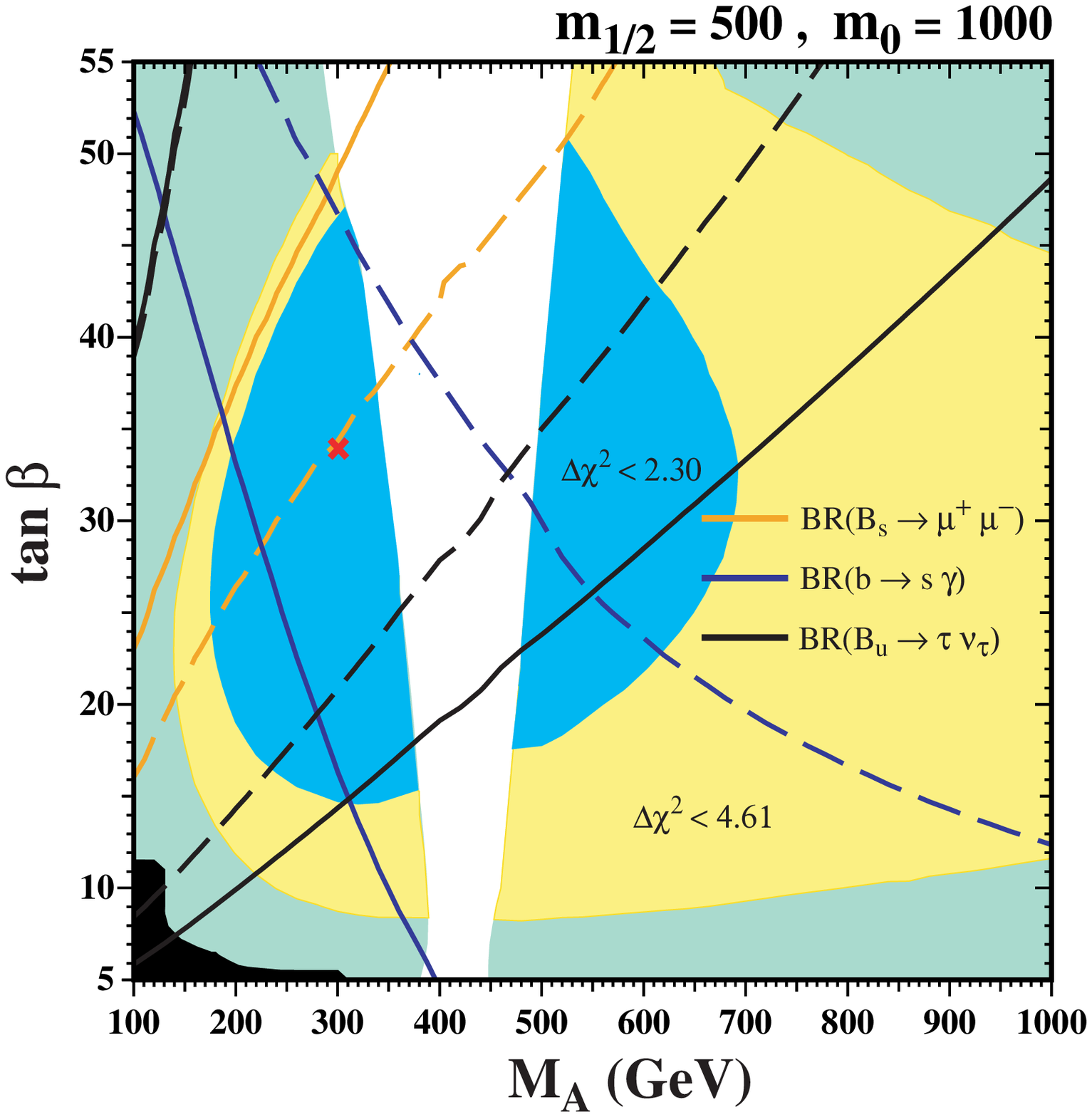}
\caption{%
\it The $(M_A, \tan \beta)$ planes for the NUHM2 benchmark surfaces
(\protect\ref{plane1}, \protect\ref{plane2}) displaying (top) the 5-$\sigma$ discovery contours 
for $H/A \to \tau^+ \tau^-$ and $H^\pm \to \tau^\pm \nu$
detection in the
CMS detector, and (bottom) the sensitivities provided by $B$ decays:
$B_s \to \mu^+ \mu^-$, $b \to s \gamma$ and $B_u \to \tau \nu$~\protect\cite{EHHOW}.}
\label{fig:EHHOW2}
\end{center}
\vspace{1em}
\end{figure}

The left panels of Fig.~\ref{fig:EHHOW2} display the plane (\ref{plane1}), and the
right panels show the plane (\ref{plane2})~\cite{EHHOW}. 
The black regions are excluded by the LEP Higgs
searches, the regions shaded blue (yellow) are favoured by the global fit at the
68\% (95\%) confidence levels, and the best-fit points in these planes are indicated
by red crosses. Also shown in the top panels are the regions accessible to LHC searches
for the heavy MSSM Higgs bosons in the channels $H/A \to \tau \tau$ and $H^\pm \to \tau \nu$.
We see that the $H^\pm$ search is likely to be effective at large $\tan \beta$ and small $m_A$,
whereas the $H/A$ searches should be able to extend to lower $\tan \beta$ and larger $m_A$,
covering (most of) the 68\% confidence-level regions.

The lower panels of Fig.~\ref{fig:EHHOW2} display the sensitivities of various $B$-physics
observables in the planes (\ref{plane1}, \ref{plane2})~\cite{EHHOW}, including $B_u \to \tau \nu$ 
(which is sensitive directly to the charged Higgs boson), $B_s \to \mu \mu$ (which is sensitive to
the closely-related heavy neutral Higgs bosons $H/A$), and $b \to s \gamma$ (which is
sensitive to cancellations between heavy Higgs contributions and sparticle exchanges).
We see that these have good prospects for exploring large areas of the $(m_A, \tan \beta)$ 
planes, including the regions favoured at the 68\% confidence level. Thus these indirect
searches play important roles in the outlook for charged Higgs physics.

\section{Charged Higgs Effects in $K$ Physics?}

It has recently been pointed out~\cite{MPP} that there is an interesting opportunity to study
charged Higgs physics in $K$ decays by probing lepton universality in
$K \to \ell \nu$ decays, via the ratio
\begin{equation}
R^{LFV}_K \; \equiv \; \frac{\Sigma_i \Gamma (K \to e \nu_i)}{\Sigma_i \Gamma (K \to \mu \nu_i)}.
\label{RLFV}
\end{equation}
The present experimental range for the new physics contribution to
$R^{LFV}_K$ is $(- 0.063, 0.017)$, and the NA62 experiment at CERN is set to reduce
this range by an order of magnitude. There are two potentially-important charged-Higgs
contributions to $R^{LFV}_K$: from lepton-flavour-violating contributions to
$\ell H^\pm \nu_\tau$ vertices:
\begin{equation}
\ell H^\pm \nu_\tau \; \to \; 
\frac{g_2}{\sqrt{2}} \frac{m_\tau}{M_W} \Delta^{3\ell}_R \tan^2 \beta, \ell = e, \mu ,
\label{LFVH}
\end{equation}
and from non-universality in the lepton-flavour-conserving $H^\pm$ vertices:
\begin{equation}
\ell H^\pm \nu_\ell \; \to \; 
\frac{g_2}{\sqrt{2}} \frac{m_\ell}{M_W} \tan^2 \beta \left( 1 + 
\frac{m_\tau}{m_\ell} \Delta^{\ell \ell}_{RL} \tan \beta \right), \ell = e, \mu .
\label{LFCH}
\end{equation}
In combination, these yield
\begin{equation}
R^{LFV}_{K} \; = \; \left[ | 1 - \frac{m_K^2}{M_H^2}
\frac{m_\tau}{m_e} \Delta^{11}_{RL} \tan^3 \beta |^2
+ \left( \frac{m_K^4}{M_H^4} \right) \left( \frac{m_\tau^2}{m_e^2} \right)
|\Delta^{31}_R|^2 \tan^6 \beta \right].
\label{combineLFV}
\end{equation}
The first term in (\ref{combineLFV}) gets contributions from mixing in both the
left- and right-handed slepton sectors, whereas the second term comes exclusively
from mixing among the spartners of right-handed leptons.

Left-handed mixing is constrained strongly by the upper limit on $\tau \to e \gamma$
decay, so the dominant combination may come from mixing in the right-handed
slepton sector~\cite{ELR}. This does not arise in a minimal SU(5) GUT, even including the
seesaw mechanism for neutrino masses. However, it may arise in non-minimal GUTs,
and make a contribution to (\ref{RLFV}) that is observable in the NA62
experiment, despite the upper limit on $\tau \to e \gamma$ decay. Maybe NA62
will find the first evidence for a charged Higgs boson?

\section{CP Violation in Higgs Physics?}

As discussed previously, upper limits on flavour violation beyond the Standard Model
suggest that the supersymmetry-breaking
scalar masses and trilinear scalar couplings are universal at high scale for all sparticles 
with the same quantum numbers, in which case the following are independent parameters:
\begin{equation}
m^2_{{\tilde Q}, {\tilde L}, {\tilde U}, {\tilde D}, {\tilde E}} \propto {\mathbf 1}_3; \; \; \;
A_{u, d, e} \propto {\mathbf 1}_3; \; \; \; M_{1, 2, 3}.
\label{MCPMFV}
\end{equation}
In this case, all squark mixing is due to the Cabibbo-Kobayashi-Maskawa (CKM) matrix,
a scenario known as minimal flavour violation (MFV). This framework has a total of 19
parameters, of which 6 violate CP~\cite{MCPMFV}, namely:
\begin{equation}
{\rm Im} A_{u, d, e}; \; \; \; {\rm Im} M_{1, 2, 3}.
\label{phases}
\end{equation}
It is often assumed that the parameters ${\rm Im} M_a$, ${\rm Im} A_f$ are
universal, but non-universality is completely compatible with MFV. The scenario with all
6 CP-violating phases left free is the maximally CP-violating, minimally flavour-violating
(MCPMFV) variant of the MSSM~\cite{MCPMFV}.

\begin{figure}
\begin{center}
\resizebox{0.8\textwidth}{!}{%
  \includegraphics{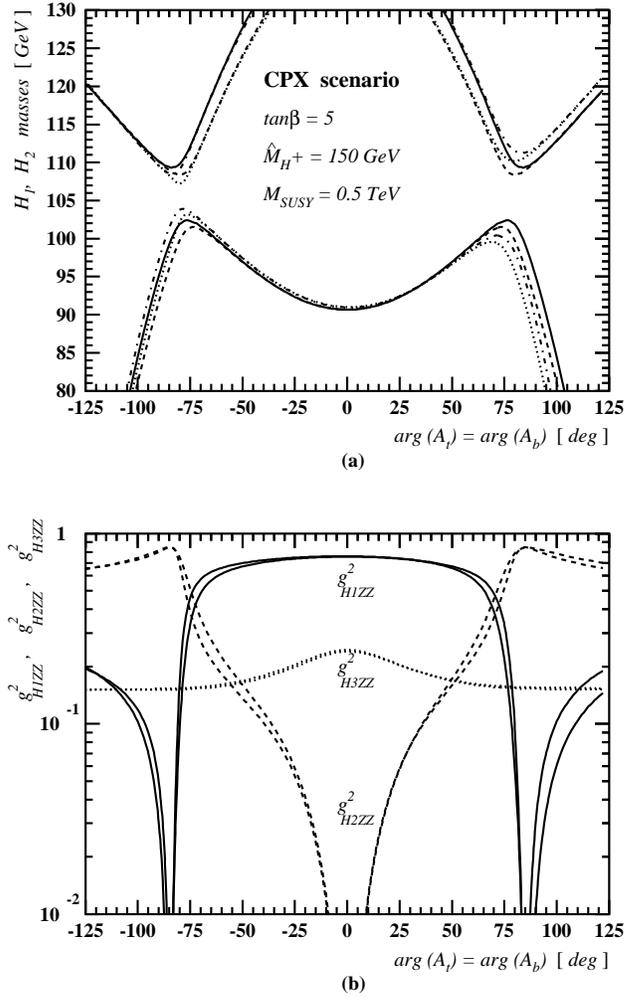}
}
\end{center}
\caption{\it The dependences on one CP-violating phase of (top) the two lightest
neutral Higgs mass eigenstates $H_{1,2}$ and (bottom) the couplings of the
three neutral Higgs bosons to pairs of $Z$ bosons, for the indicated values of
other supersymmetric parameters~\protect\cite{CEPW}.
}
\label{fig:CEPW}       
\end{figure}

In the presence of CP violation, there is mixing between the scalar Higgs bosons
$h, H$ and the pseudoscalar $A$, and one may label the mass eigenstates
$H_{1, 2, 3}$ in order of increasing mass~\cite{CPsuperH}. 
In this case, since the pseudoscalar state is no longer
a mass eigenstate, it is better to use the charged Higgs mass as a parameter
to characterize the MSSM Higgs sector, e.g., by displaying $(m_{H^\pm}, \tan \beta)$
planes instead of $(m_A, \tan \beta)$ planes. Fig.~\ref{fig:CEPW} shows some effects of
CP-violating phases on the masses and couplings of the MSSM Higgs bosons~\cite{CEPW},
under the simplifying assumption that the phases ${\rm Im} A_{u, d, e}$ and ${\rm Im} M_{1, 2, 3}$
are each universal. We see interesting level-crossing phenomena in the masses, and possible
suppressions of the $H_i V V$ couplings, which could have important consequences for
phenomenology. For example, the LEP exclusion of a neutral Higgs below 114~GeV
would no longer be valid in a CP-violating scenario, and there would be new challenges for
Higgs searches at the LHC, because of the different pattern of Higgs couplings~\cite{CEMPW}.

The CP-violating phases in the MCPMFV are strongly constrained by the experimental
upper limits on electric dipole moments (EDMs), notably those of Thallium, the neutron
and Mercury~\cite{Pospelov}. However, there is the possibility of non-trivial cancellations between the 
contributions of different phases~\cite{ELPEDM}. The three EDM constraints cannot force all the 6
CP-violating phases of the MCPMFV model to be small simultaneously.

\begin{figure}[ht]
\hspace{ 0.0cm}
\centerline{\epsfig{figure=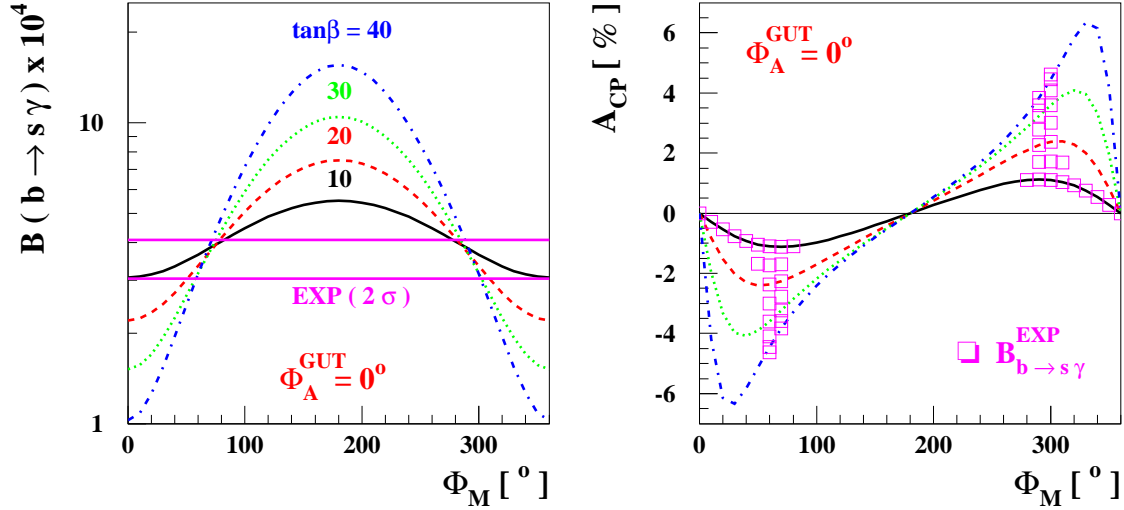,height=16cm,width=16cm}}
\vspace{-7.9cm}
\caption{\it The branching ratio  $B(B \rightarrow X_s \gamma)$ (left)
and the  CP asymmetry ${\cal  A}^{\rm dir}_{\rm CP}(B  \rightarrow X_s
\gamma)$  (right)  as  functions   of  the common gaugino mass phase $\Phi_M$  for  four  values  of
$\tan\beta$, taking the trilinear coupling phase $\Phi_A^{\rm GUT}=0^{\rm o}$.   The region allowed
experimentally at the 2-$\sigma$  level is bounded by two horizontal
lines in the  left frame.  In the right  frame, points satisfying this
constraint are denoted  by open squares: see~\protect\cite{MCPMFV} for details. }
\label{fig:ELPB}
\end{figure}

The MCPMFV phases may have important implications for the heavy Higgs
contributions to $B$-physics observables, such as $B_s$ mixing, $B_s \to \mu \mu$,
$B_u \to \tau \nu$ and $b \to s \gamma$~\cite{MCPMFV}. Even after taking into account the
constraints these observables impose, it is possible that there might be an important
contribution from MCPMFV phases to CP-violating observables such as the rate and
CP-violating asymmetry in $b \to s \gamma$ decay, as seen in Fig.~\ref{fig:ELPB}.

One should be on the lookout not just for CP-conserving effects of charged Higgs
bosons, but also for the possibility that they may manifest CP violation beyond the CKM
model.

\section{Summary}

The existence of charged Higgs bosons is generic in physics beyond the Standard Model.
In this talk I have focused on the MSSM s one particularly predictive example, but similar
considerations apply to many other proposed extensions of the Standard Model.

The LHC offers real prospects for direct detection of charged Higgs bosons,
and is coming on line. After its brilliant startup and the subsequent
disappointment, we are all crossing our fingers for high-energy collisions in 2009.

However, the LHC is not the only game in town: there are also opportunities to
probe charged Higgs bosons indirectly in low-energy physics. Examples include
tests of lepton universality in $K \to \ell \nu$ decays, a detectable contribution to
$B_u \to \tau \nu$ decay, CP-violating effects, etc.

Floreat charged Higgs physics!

\end{document}